\documentclass[]{aipproc}
\layoutstyle{6x9}

\SetInternalRegister\hbadness{8000} 

\usepackage{epsfig}
\usepackage{here}
\usepackage[scriptsize]{subfigure} 

\begin{document}

\title 
      [Production of $\eta$ and $\eta^{\prime}$ mesons]
      {Production of $\eta$ and $\eta^{\prime}$ mesons via the quasi-free 
             proton-neutron interaction}


\newcommand{\ikpjuel}{IKP, Forschungszentrum J\"{u}lich, D-52425 J\"{u}lich, Germany}
\newcommand{\ikpmue}{IKP, Westf\"{a}lische Wilhelms--Universit\"{a}t, D-48149 M\"{u}nster, Germany}
\newcommand{\cracow}{M.~Smoluchowski Institute of Physics, Jagellonian University, PL-30-059 Cracow, Poland}
\newcommand{\nphycracow}{Institute of Nuclear Physics, PL-31-342 Cracow, Poland}
\newcommand{\uppsala}{Uppsala University, S-75121 Uppsala, Sweden}
\newcommand{\catowice}{Institute of Physics, University of Silesia, PL-40-007 Katowice, Poland}
\newcommand{\zeljuel}{ZEL,  Forschungszentrum J\"{u}lich, D-52425 J\"{u}lich,  Germany}

\author{P.~Moskal \thanks{email: \tt{p.moskal@fz-juelich.de}}\ }{address={\ikpjuel}}
\author{H.-H.~Adam}{address={\ikpmue}}
\author{A.~Budzanowski}{address={\nphycracow}}
\author{R.~Czy{\.{z}}ykiewicz}{address={\ikpjuel}}
\author{D.~Grzonka}{address={\ikpjuel}}
\author{M.~Janusz}{address={\cracow}}
\author{L.~Jarczyk}{address={\cracow}}
\author{T.~Johansson}{address={\uppsala}}
\author{B.~Kamys}{address={\cracow}}
\author{A.~Khoukaz}{address={\ikpmue}}
\author{K.~Kilian}{address={\ikpjuel}}
\author{P.~Kowina}{address={\ikpjuel}}
\author{W.~Oelert}{address={\ikpjuel}}
\author{C.~Piskor-Ignatowicz}{address={\cracow}}
\author{J.~Przerwa}{address={\cracow}}
\author{T.~Ro{\.{z}}ek}{address={\ikpjuel}}
\author{R.~Santo}{address={\ikpmue}}
\author{G.~Schepers}{address={\ikpjuel}}
\author{T.~Sefzick}{address={\ikpjuel}}
\author{M.~Siemaszko}{address={\cracow}}
\author{J.~Smyrski}{address={\cracow}}
\author{A.~Strza{\l}kowski}{address={\cracow}}
\author{A.~T\"aschner}{address={\ikpmue}}
\author{P.~Winter}{address={\ikpjuel}}
\author{M.~Wolke\footnote{present address: The Svedberg Laboratory, Thumbergsv\r{a}gen 5A, Box 533, S-75121 Uppsala, Sweden.}\ }{address={\ikpjuel}}
\author{P.~W{\"u}stner}{address={\zeljuel}}
\author{W.~Zipper}{address={\catowice}}

\copyrightyear  {2001}

\begin{abstract}
A comparison of the close-to-threshold total cross sections for the $\eta^{\prime}$
meson production in both the $pp\to pp\eta^{\prime}$ and $pn \to pn \eta^{\prime}$ reactions 
should provide  insight into the flavour-singlet (perhaps also into gluonium)
content of the $\eta^{\prime}$ meson and the relevance of quark-gluon
or hadronic degrees of freedom in the creation process.
  The excitation function
  for the reaction $pp\to pp\eta^{\prime}$
  has been already established.
At present, experimental investigations of the quasi-free $pn \to pn X$ 
reactions are carried out at the COSY-11 facility using a beam of 
stochastically cooled protons and the deuteron
cluster target. 
A method of measurement and preliminary results from the test experiments of the $pn\to pn \eta$ reaction are presented
in this report. 
\end{abstract}

\date{\today}

\maketitle

\section{Introduction}

Close-to-threshold production of $\eta$ and $\eta^{\prime}$ mesons
in the nucleon-nucleon interaction requires a large momentum 
transfer between the nucleons and occur at distances
in the order of $\sim$0.3~fm.  
This implies that the quark-gluon 
degrees of freedom may 
play a significant role in the production dynamics of these mesons.
Therefore, additionally to the mechanisms associated with meson
exchanges it is possible that the $\eta^{\prime}$ meson is created from excited glue
in the interaction region of the colliding nucleons~\cite{bass99,bass348}, 
which couple to the $\eta^{\prime}$ meson directly via its gluonic
component or through its SU(3)-flavour-singlet admixture. The production through the
colour-singlet object as suggested in reference~\cite{bass99} is isospin independent
and should lead to the same production yield of the
$\eta^{\prime}$ meson in the $pn\to pn\eta^{\prime}$ and $pp\to pp\eta^{\prime}$ reactions
after correcting for the final and initial state interaction between the nucleons.
 
Investigations of the $\eta$-meson production  in collisions of nucleons
allowed to conclude that, close to the kinematical threshold,
the creation of $\eta$ meson from isospin I~=~0 
exceeds the production with I~=~1
by about a factor of 12. This  was derived from the measured ratio of the 
total cross sections for the reactions $pn \to pn \eta$ and 
$pp \to pp\eta$ ($R_{\eta} = \frac{\sigma{(pn\to pn\eta})}{\sigma{(pp\to pp\eta)}}$), 
which was determined to be $R_{\eta}\approx 6.5$ in the excess energy 
range between 16~MeV and 109~MeV~\cite{calen_pneta}.
The large difference of the total cross section
between the isospin channels
suggests the dominance of
isovector meson ($\pi$ and $\rho$) exchange
in the creation of $\eta$ in nucleon-nucleon collisions~\cite{wilk,calen_pneta}.

Since the quark structure of $\eta$ and $\eta^{\prime}$ mesons is very similar
we can --~by analogy to the $\eta$ meson production~-- 
expect that in the case of  
dominant isovector meson exchange the ratio $R_{\eta^{\prime}}$  
should also be about 6.5. If, however, the $\eta^{\prime}$ meson was produced via its 
flavour-blind gluonic component from the colour--singlet glue excited
in the interaction region, the ratio $R_{\eta^{\prime}}$ should approach unity
after corrections for the interactions between the participating baryons.

\begin{figure}[H]
  {\epsfig{file=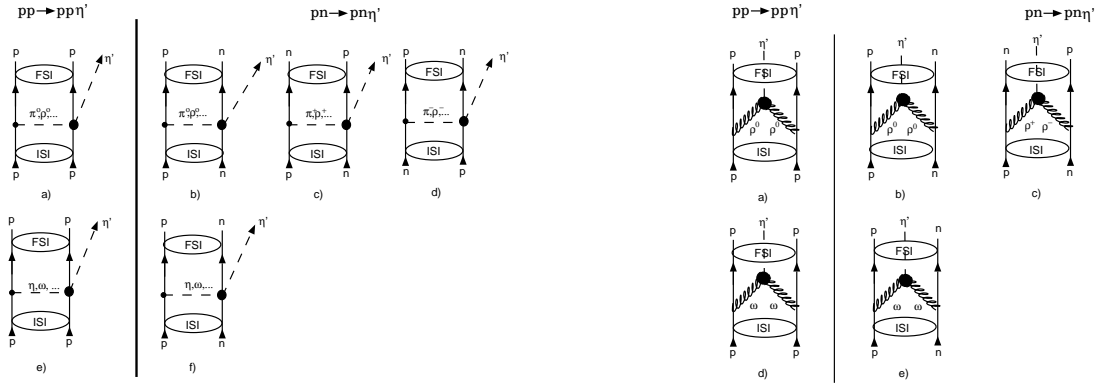,width=0.35\textwidth,angle=270}}  
  {\caption{\label{graf_isospin1} ({\bf left}) Example of diagrams with the isovector
  and isoscalar meson exchange leading to the creation of the meson
  $\eta^{\prime}$ in the proton-proton and proton-neutron collisions.
  ({\bf right})
  Fusion of the virtual $\omega$ and $\rho$ mesons
  emitted from the colliding nucleons. }}
\end{figure}

Figure~\ref{graf_isospin1} demonstrates  
qualitatively the fact that the production of mesons in the proton-neutron
collisions is more probable than in the proton-proton interaction if it
is driven by the isovector meson exchanges only. 
This is because in the case of the proton-neutron collisions there are always more possibilities
to realise the exchange or fusion of the
isovector mesons than in the case of the reaction of protons. 

The close-to-threshold 
excitation function for the $pp\to pp\eta^{\prime}$ 
reaction has been already determined~\cite{moskalprl,moskalpl,hiboupl,disto_etap}
whereas the total cross section for 
the $\eta^{\prime}$ meson production in the proton-neutron 
interaction remains unknown. As a first step towards the determination of the value of $R_{\eta^{\prime}}$
the feasibility of the measurement of the $pn\to pn\eta^{\prime}$ reaction by means of the COSY-11 facility
was studied using a Monte-Carlo method~\cite{moskal0110001,rafalmgr}. As a second step, 
a test experiment of the $pn\to pn\eta$ reaction --~suspected  to have by at least a factor of  30 larger cross section than the one
for the $pn\to pn\eta^{\prime}$ reaction~--
was performed.
  In this test measurement,
  using a beam of protons and a deuteron cluster target,
  we have proven the ability of the COSY-11 facility to study 
  the quasi-free creation of mesons via the $pn\to pn X$ reaction.
Appraisals of simulations and preliminary  results of the measurements of the quasi-free $pn\to pn\eta$
reaction performed using the newly extended COSY-11 facility~\cite{moskal0110001,rafalmgr} 
will be presented in the next section.

\section{Test measurement of the $pn \to pn \eta$ reaction }
\begin{figure}[H]
\parbox{0.55\textwidth}{\epsfig{file=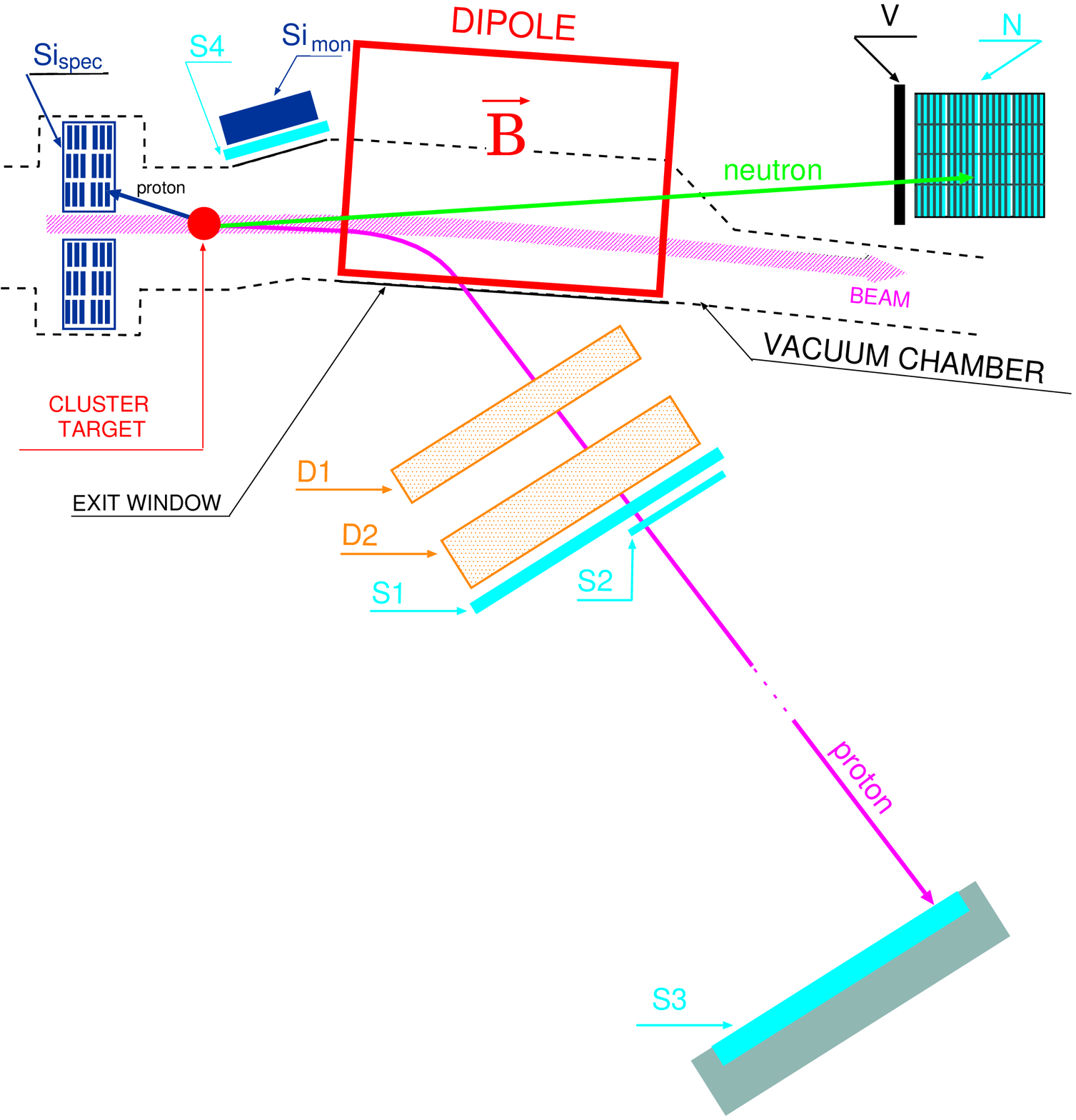,width=0.55\textwidth,angle=0}}
\parbox{0.44\textwidth}{
            \caption{\small  Schematic view of the COSY-11 detection setup~\protect\cite{brau96}.
             Only detectors needed for the measurements of the  reaction
             $pd\rightarrow p_{sp}pn\eta(\eta^{\prime})$
             are shown. \protect\\
             D1, D2 denote the drift chambers; S1, S2, S3, S4 and V the scintillation detectors;
             N the neutron detector and Si$_{mon}$ and Si$_{spec}$ silicon strip detectors
             to detect elastically scattered and spectator protons, respectively.
             \label{detectionsystem}
            }
          }
\end{figure}

\begin{figure}[t]
       \vspace{-2.0cm}
       \hspace{-1.7cm} 
       \epsfig{file=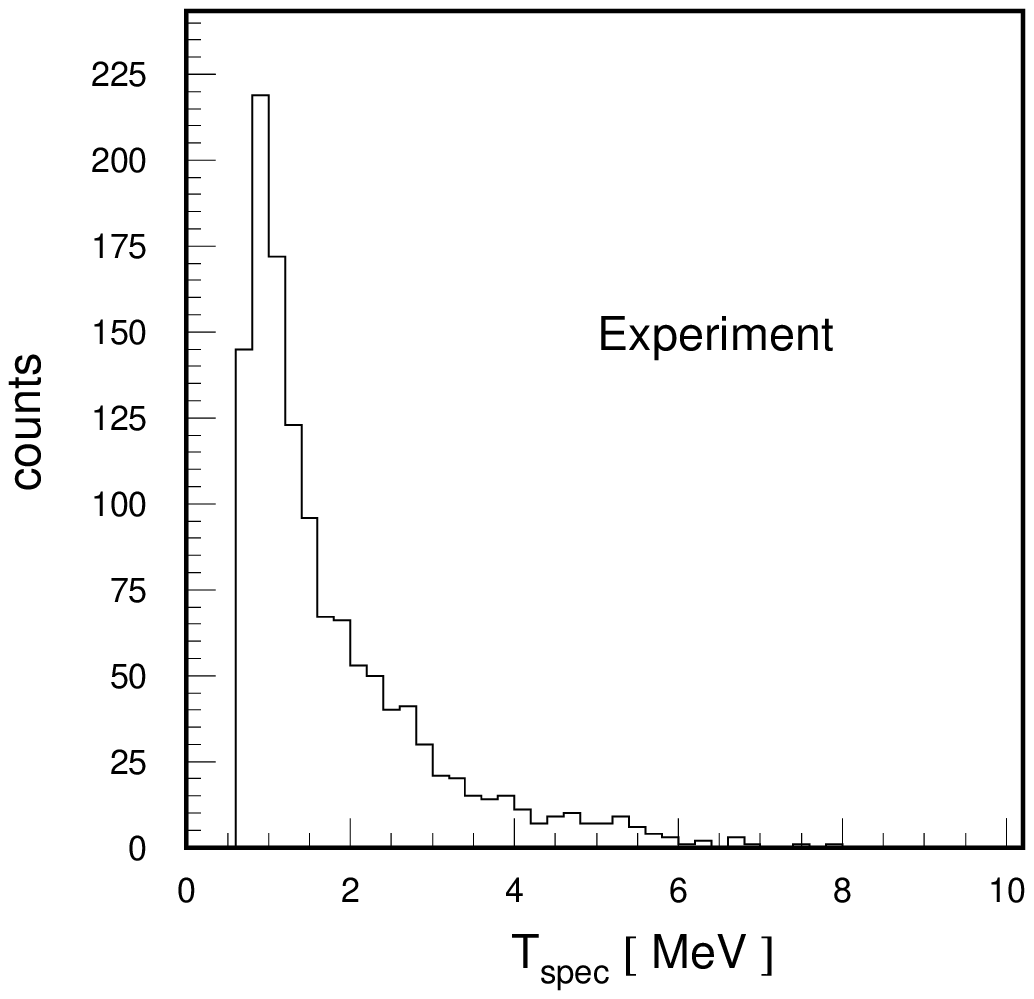,width=0.48\textwidth,angle=0}
       \epsfig{file=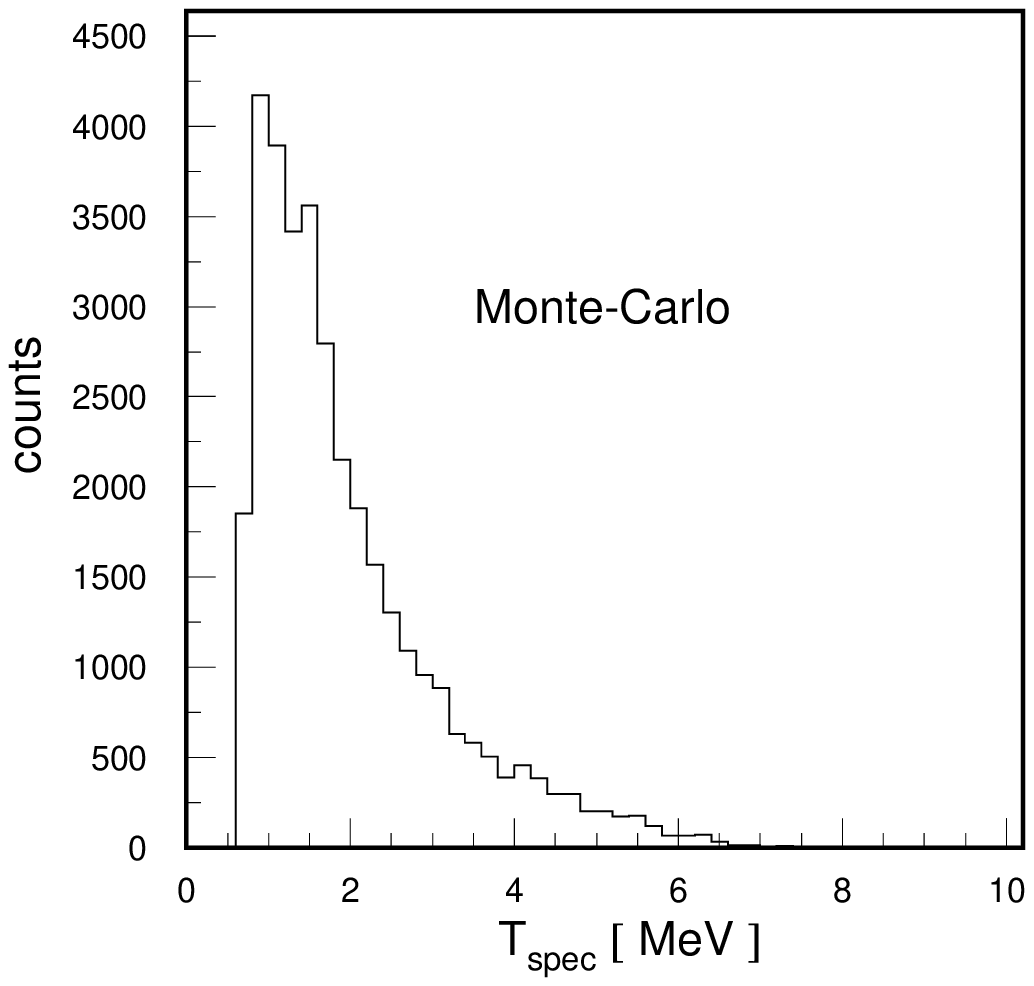,width=0.48\textwidth,angle=0}\\
       \hspace{-1.0\textwidth}
       \parbox{0.45\textwidth}{\mbox{} }\hfill
       \parbox{0.51\textwidth}{ \vspace{-0.7cm} a) }\hfill
       \parbox{0.04\textwidth}{ \vspace{-0.7cm} b) }\\
   \caption{ \small
          Distributions of the kinetic energy of the spectator protons.\protect\\
         (a) Experiment, \ \ \
         (b) Monte-Carlo simulations taking into account the acceptance
             of the COSY-11 detection system and 
           an analytical parametrization of the
           deuteron wave function~\protect\cite{lacombe81}
           calculated from the PARIS potential~\protect\cite{lacombe80}.
             \label{exp1}
         }
\end{figure}

As a general commissioning of the extended COSY-11 facility
to investigate quasi-free $pn\to pn X$ reactions,
we have performed a measurement of the $pn \to pn\eta$ process
at a beam momentum of 2.075~GeV/c. The experiment, carried out
in June 2002,
had been preceded  by the  installation of a spectator~\cite{bilger} and  neutron detectors,
and by a series of thorough simulations performed
in order to determine the best conditions for measuring quasi-free
$pn\to pn \eta$ and $pn\to pn \eta^{\prime}$ reactions~\cite{moskal0110001,rafalmgr}.
Figure~\ref{detectionsystem} presents the COSY-11 detection facility with  superimposed tracks
of protons and neutron originating from the quasi-free $pn\to pn X$ reaction induced  
by a proton beam~\cite{cosy_dieter} impinging on a deuteron target~\cite{domb97}. 
The  identification of the $pn\to pn\eta$ reaction is based on the measurement of the four-momentum vectors
of the outgoing nucleons and the $\eta$ meson is identified via the missing mass technique.
The slow proton  stopped in the first layer of the position sensitive silicon detector (Si$_{spec}$)
is, in the analysis, considered 
as a spectator without interaction
with the bombarding particle 
and it is moving with the  Fermi momentum as possessed 
at the moment of the collision.
From the measurement of the momentum vector of the spectator proton
$\vec{\mbox{p}}_{sp}$ one can infer the momentum vector of the struck neutron
$\vec{\mbox{p}}_n = - \vec{\mbox{p}}_{sp}$ at the time of the reaction and hence
calculate the total energy of the colliding nucleons for each event.
In the  approximation that the 
struck neutron is treated as a free particle
we can assume that the matrix element for quasi--free meson production off a
bound neutron is identical to that for the free $pn \rightarrow p n\,Meson$
reaction~\footnote{For more comprehensive discussion of this issue the reader is referred to reference~\cite{review}}.
In figures~\ref{exp1}a and~\ref{exp1}b the measured and expected distribution of the kinetic energy of the spectator proton 
is presented. Though still very rough energy calibration of the detector units  one recognizes a substantial similarity in the shape 
of both distributions. 

\vspace{-1.2cm}

  \begin{figure}[h]
   \caption{ \small
          Distributions of the excess energy $Q_{CM}$
           for the quasi-free $pn\rightarrow pnX$ reaction,
           determined with respect to the $pn\eta$ threshold.
           (a) Experiment. (b) Simulation.
             \label{exp2}
         }
       \epsfig{file=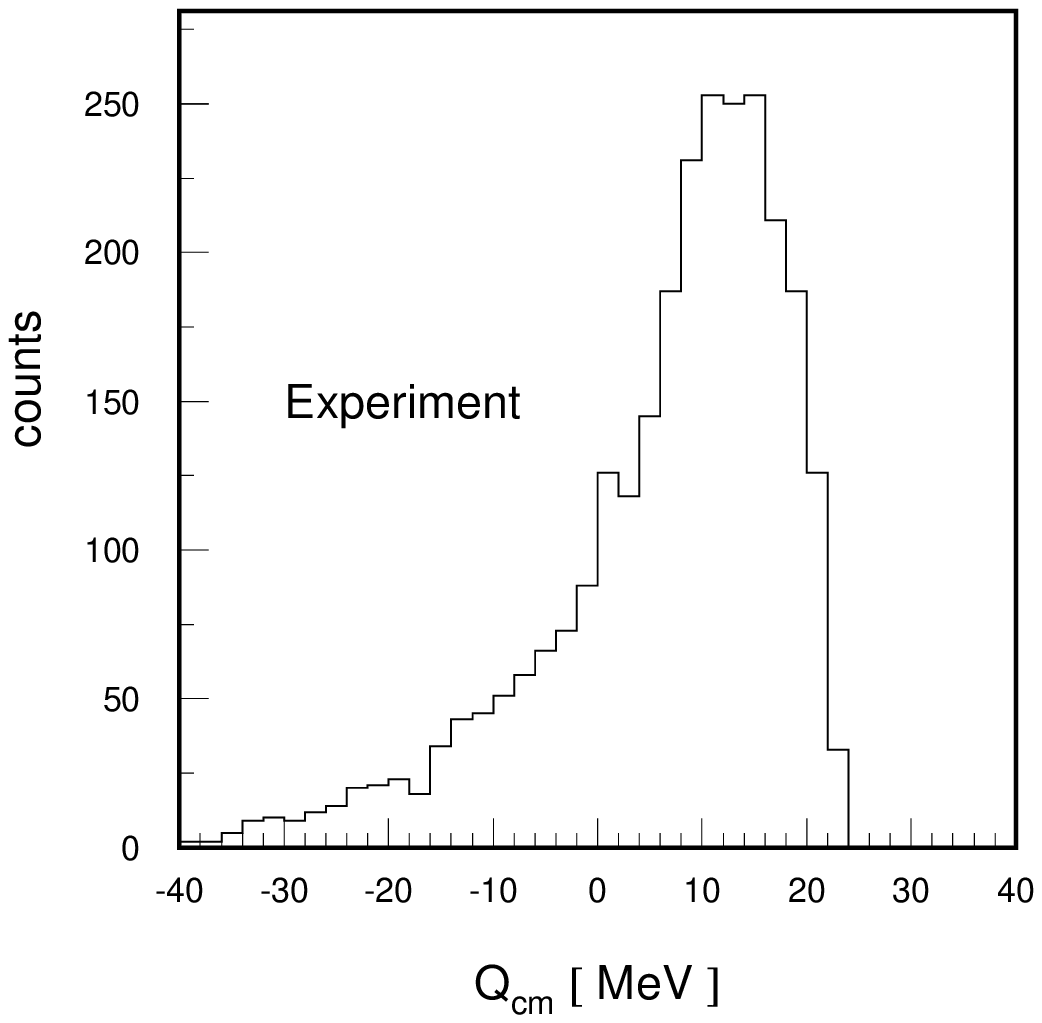,width=0.48\textwidth,angle=0}
       \epsfig{file=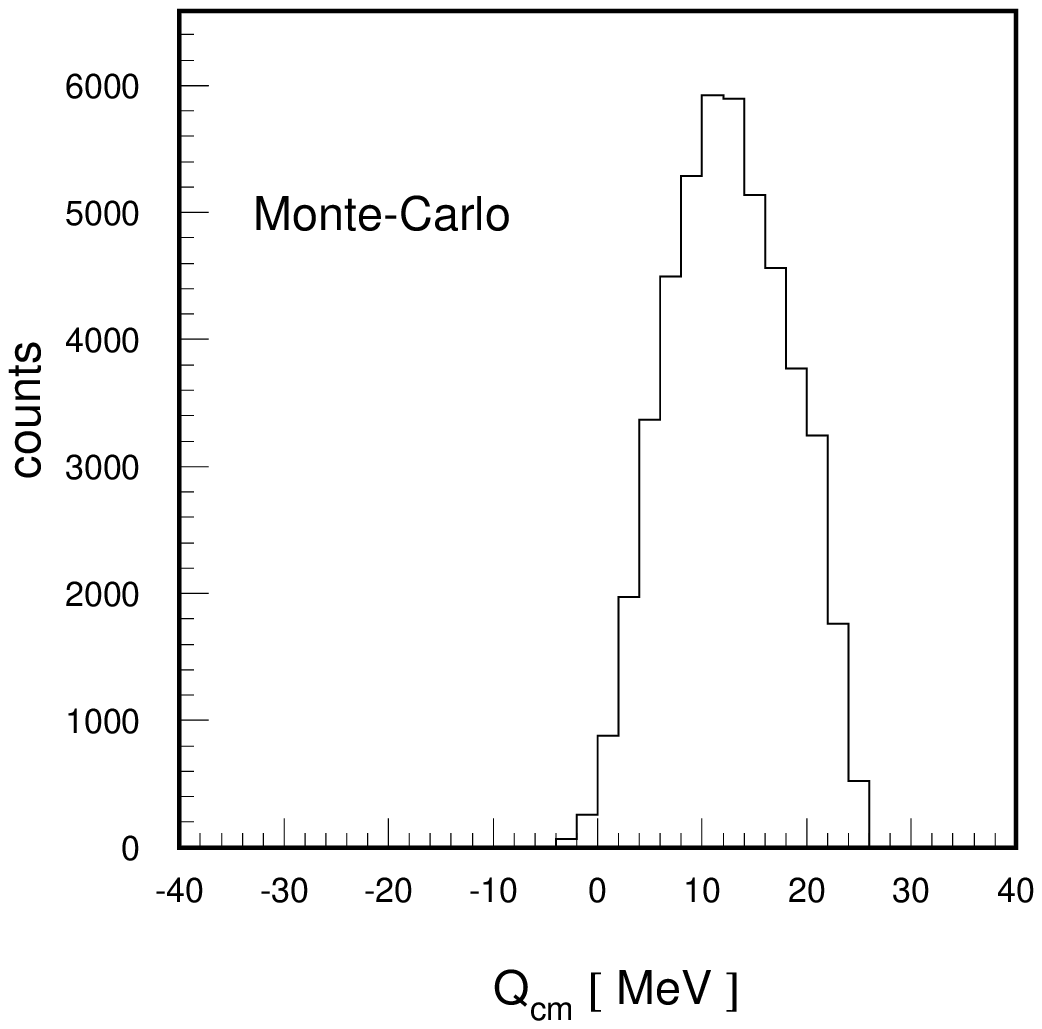,width=0.48\textwidth,angle=0}
       \hspace{-1.0\textwidth}
       \parbox{0.45\textwidth}{\mbox{} }\hfill
       \parbox{0.51\textwidth}{ \vspace{-0.7cm} a) }\hfill
       \parbox{0.04\textwidth}{ \vspace{-0.7cm} b) }
\end{figure}
\begin{figure}[t]
       \epsfig{file=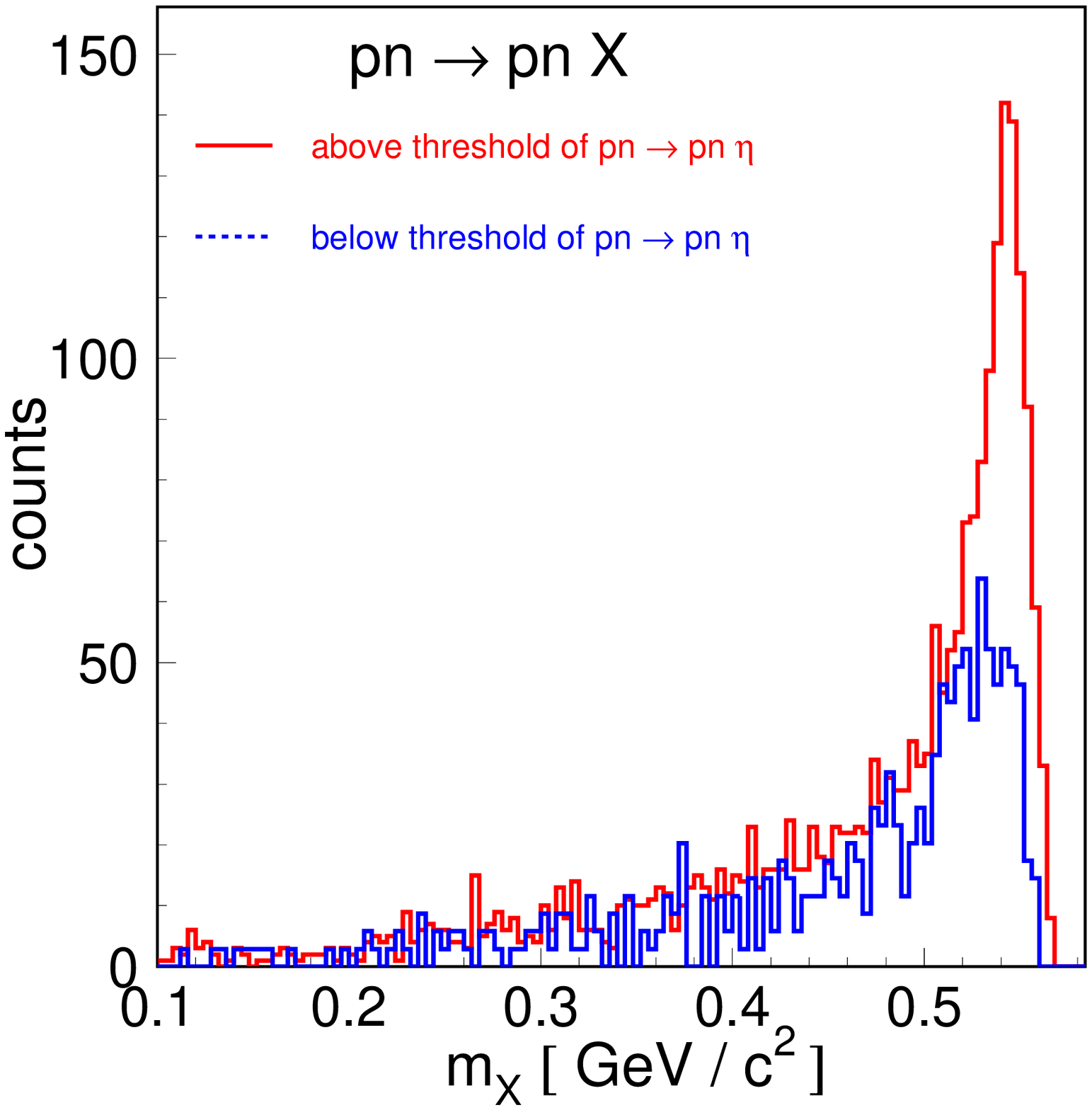,width=0.42\textwidth,angle=0}
       \epsfig{file=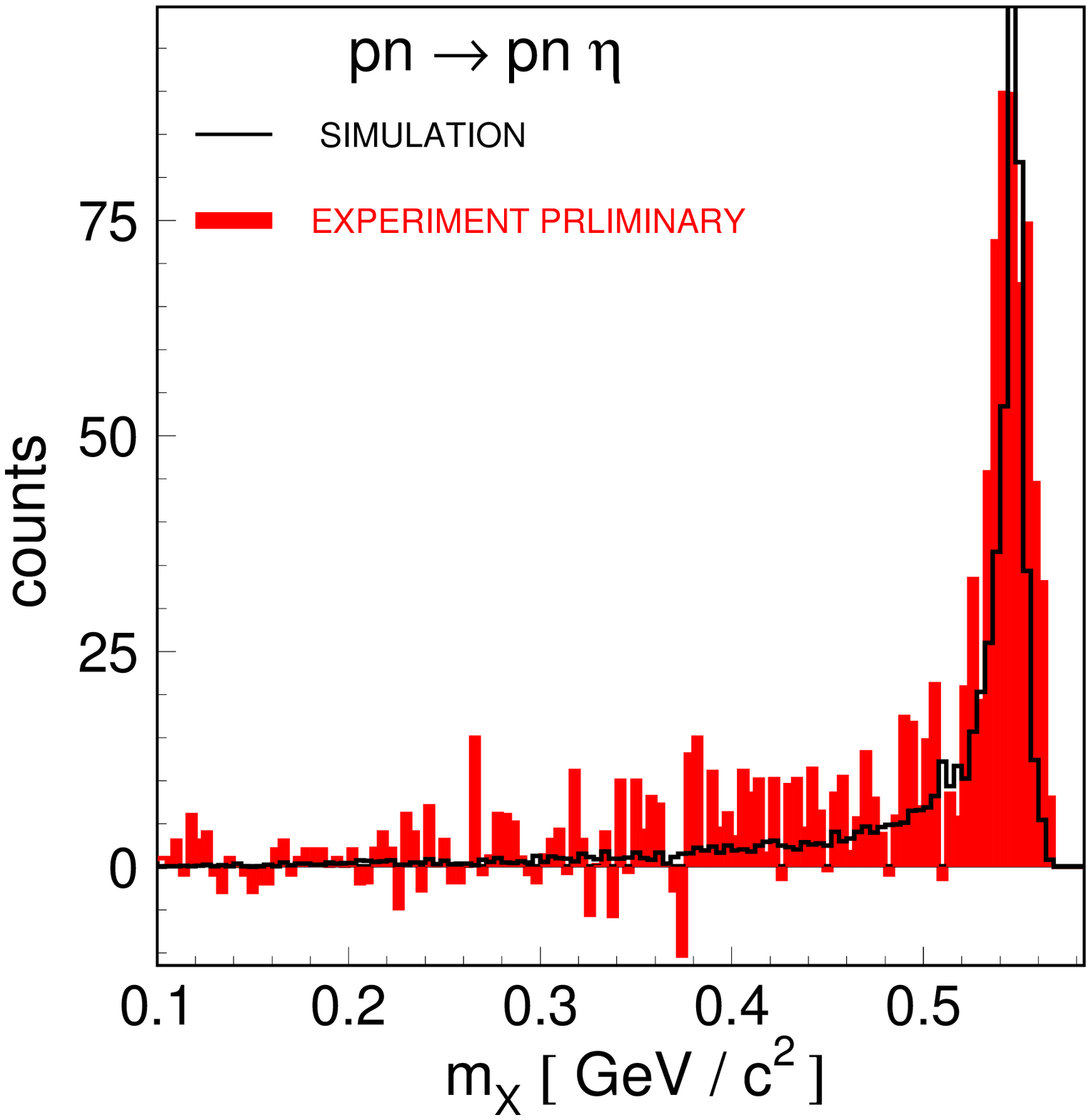,width=0.42\textwidth,angle=0}\\
       \hspace{-1.0\textwidth}
       \parbox{0.49\textwidth}{\mbox{} }\hfill
       \parbox{0.46\textwidth}{ \vspace{-0.7cm} a) }\hfill
       \parbox{0.01\textwidth}{ \vspace{-0.7cm} b) }
   \caption{ \small
          Missing mass spectra as obtained during the June'02 run: \protect\\
          a) Event distribution for $Q < 0$ ( black line) and for $Q > 0$ (gray line).\protect\\
          b) Histogram represents the difference between number of events above and below threshold
             for the $pn\to pn\eta$ reaction, and the line corresponds to the Monte-Carlo simulation.
          \label{exp6}
         }
\end{figure}

Figure~\ref{exp2} shows spectra of the excess energy in respect to the $pn\eta$ system
as obtained in the experiment~(\ref{exp2}a) and
the simulation~(\ref{exp2}b) for the $pn \rightarrow p n\eta$ reaction. The remarkable difference between the distributions comes from the
fact that in reality additionally to the $pn\to pn\eta$ reaction also the multi-pion production is registered.
The $\eta$ and multi-pion production cannot be distinguished from each other on the event-by-event basis by means of the missing mass technique.
However, we can determine the number of the registered $pn\to pn\eta$ reactions from the multi-pion background
comparing the missing mass distributions for Q values larger and smaller than zero. Knowing that negative values of Q can only be
assigned to the multi-pion events we can derive the shape of the missing mass ditribution corresponding to these events.
This is shown as the solid line in figure~\ref{exp6}.  A thorough evaluation of the background is in progress, however, rough 
comparison of events for positive and negative Q yields the promissing results with a clear signal from the  $pn\to pn\eta$ reactions,
as can be deduced by inspection of figures~\ref{exp6}a and ~\ref{exp6}b.



\bibliographystyle{aipproc}
\bibliography{abbrev,general}

\end{document}